%% file: main.tex
\documentclass[letterpaper, 10 pt, conference]{ieeeconf}  % Comment this line out
\IEEEoverridecommandlockouts
\usepackage{cite}
\usepackage{nicefrac}

\usepackage{ulem}
\usepackage{xcolor}
\usepackage{multirow, lipsum}
\usepackage{enumerate}
\usepackage{booktabs} % For formal tables
\usepackage{amsmath}
\usepackage{physics}
\usepackage{dingbat}
\usepackage{hyperref}
\usepackage{url}
\usepackage{graphicx, subcaption}
\usepackage{siunitx}
\usepackage{import}
\usepackage{placeins}
\usepackage{amsthm}
\usepackage{amssymb}
\usepackage{empheq}
\usepackage{ulem}
\usepackage{algorithm}
\usepackage{float}
\usepackage{multicol}
\usepackage{cleveref}
\usepackage[english]{babel}
\usepackage{pifont}

\newcommand{\wrong}{\ding{55}}
\newcommand{\hide}[1]{}

\newcommand{\eop}{\hfill{$\square$}}

\newtheorem*{example}{Example}
\newtheorem{theorem}{Theorem}
\newtheorem{proposition}{Proposition}
\newtheorem{corollary}{Corollary}
\newtheorem{definition}{Definition}

\newcommand{\supp}{\mbox{support}}
\newcommand{\us}{u^S}
\newcommand{\ur}{u^R_{1}}
\newcommand{\ura}{u^R_{\alpha}}
\newcommand{\uira}{u^{I}_{\alpha}}
\newcommand{\poar}{\rho^{R}_{\alpha}}
\newcommand{\poair}{\rho^{I}_{\alpha}}
\newcommand{\poa}{\rho_{1}}
\newcommand{\NE}{\mathcal{N}_{\alpha}}
\newcommand{\zh}{\mathcal{Z}^h}
\newcommand{\zha}{\mathcal{Z}^h_\alpha}

\newcommand{\NEc}{\mathcal{N}_{1}}

\title{Balancing rationality and social influence: \\ Alpha-rational Nash 
equilibrium in games with herding 
}

\author{Khushboo Agarwal, Konstantin Avrachenkov, Veeraruna Kavitha and Raghupati Vyas%
\thanks{K. Agarwal, V. Kavitha and R. Vyas are with Industrial Engineering and Operations Research, IIT Bombay, Powai, Mumbai, 400076, India
        {\tt\small \{agarwal.khushboo, vkavitha, raghupati.vyas\}@iitb.ac.in}}%
\thanks{K. Avrachenkov is with Inria Sophia Antipolis, 2004 Route des Lucioles, Valbonne 06902, France
        {\tt\small k.avrachenkov@inria.fr}}%
}

\begin{document}
\maketitle
\thispagestyle{empty}
\pagestyle{empty}

\begin{abstract} 
The classical game theory models rational players and proposes Nash equilibrium (NE) as the solution. However, real-world scenarios rarely feature rational players; instead, players make inconsistent and irrational decisions. Often, irrational players exhibit herding behaviour by simply following the majority.

In this paper, we consider the mean-field game with $\alpha$-fraction of rational players and the rest being herding-irrational players. For such a game, we introduce a novel concept of equilibrium named $\alpha$-Rational NE (in short, $\alpha$-RNE). The $\alpha$-RNEs and their implications are extensively analyzed in the game with two actions. Due to herding-irrational players, new equilibria may arise, and some classical NEs may be deleted. 

The rational players are not harmed but benefit from the presence of irrational players. Notably, we demonstrate through examples that rational players leverage upon the herding behaviour of irrational players and may attain higher utility (under $\alpha$-RNE) than social optimal utility (in the classical setting). 

Interestingly, the irrational players may also benefit by not being rational. We observe that irrational players do not lose compared to some classical NEs for participation and bandwidth sharing games. More importantly, in bandwidth sharing game, irrational players receive utility that approaches the social optimal utility. Such examples indicate that it may sometimes be `rational' to be irrational. 

% a game called ``$\alpha$-Rational Stackelberg mean-field game with herding" that accommodates both rational players and irrational players who exhibit herding behavior. The solution of this new game leads to a novel equilibrium concept named as $\alpha$-Rational NE ($\alpha$-RNE). 
\end{abstract}

\textbf{Keywords:} rationality, irrationality, herding, behavioural aspects, game theory, mean-field games

\section{Introduction}
Classical game theory explores the interactions between rational and intelligent players. In \cite{narahari2014game}, a player is defined as rational if it consistently makes decisions aligned with its objectives, striving to maximize its utility. Additionally, an intelligent player possesses complete knowledge of the game and can perform computations to identify its optimal strategy. This line of thought is widely acknowledged and serves as a benchmark for analyzing an ideal world. 

Nevertheless, contemporary perspectives challenge the strong assumptions regarding rationality and intelligence due to human irrationality and computational limitations. This has sparked interest in understanding actual human behavior, leading to the emergence of fields like behavioural game theory, behavioural economics, and neuro-economics (see \cite{camerer2011behavioral, thaler2018cashews, schultz2008introduction} respectively). Social experiments play a major role in driving research in these domains. 

\textit{Our work bridges classical and behavioural game theories by designing an analytical model to capture interactions between rational and irrational players}. In particular, we consider two types of players: (i) rational players, constituting an $\alpha$-fraction of the population, and (ii) \textit{irrational players, who exhibit herding behaviour}. The latter group does not optimize the utility function --- they simply choose the action taken by the majority. We specifically focus on the pervasive herding tendency among irrational players, evident in career and fashion choices driven by trends, and consumers favouring famous shops, etc. (see \cite{banerjee1992simple} for more examples).

% Following the footsteps of classical game theory, \textit{we formulate a Stackelberg game involving rational players} (at the higher level) \textit{and irrational players} (at the lower level). Assuming a large number of players, we analyze the game within a mean-field framework, yielding the `\textit{$\alpha$-Rational Stackelberg mean-field game with herding}'.
Assuming a large number of players, we analyze the game within a mean-field framework.
Drawing inspiration from the Nash equilibrium (NE) in the classical mean-field games, \textit{we introduce a novel equilibrium concept termed `\textit{$\alpha$-Rational NE}' (in short, `$\alpha$-RNE') for our game, which encompasses both the rational and irrational choices of the players}.

At this point, it is important to note that our equilibrium notion is completely different than the $k$-fault tolerant equilibrium defined in \cite{eliaz2002fault} (and subsequently analyzed in \cite{vasal2020alpha, vasal2020fault} and others). In the said literature, faulty players who deviate from the optimal behavior are considered. Their notion provides stability to the rational players against deviations by any subset of the faulty players. However, they focus on simultaneous play by the players, whereas our irrational players make their decisions only after the rational players choose their actions. Thus, the equilibrium notion in \cite{eliaz2002fault} and others can not capture the herding behavior.

Next, we conduct a more elaborate study of the game with only two actions. We \textit{establish simple conditions for identifying $\alpha$-RNEs}. It is demonstrated that the equilibrium set remains identical to the classical scenario ($\alpha = 1$) when rational players dominate the system ($\alpha > \nicefrac{1}{2}$). However, in the context of a more realistic scenario with a higher proportion of players exhibiting herding behavior ($\alpha \leq \nicefrac{1}{2}$), \textit{some classical NEs may be eliminated, while two new equilibria ($\alpha, 1-\alpha$) may emerge}. 

It is proved that rational players always gain more utility than herding-irrational players. One may think that the presence of irrational players would hurt the rational players. But, we demonstrate through various interesting examples that the contrary holds.  Interestingly, under $\alpha$-RNE, \textit{the rational players can receive more utility than the social optimal utility obtained under $\alpha=1$ case}; while, most of the time, players receive lesser utility than the social optimal utility under classical NE. 
% To see this, recall that the rational players always attain equal or lesser utility at NE than the social optimal utility. 

The incompetent behaviour of the irrational players never allows them to surpass the social optimal utility. Nonetheless, in some cases, the irrational players get utility at par with the rational players. Surprisingly, 
in an example, we observed that \textit{all the players (including irrational players) receive higher utility at $\alpha$-RNE than they could have received if everyone were rational}. Such instances encourage us to claim that \textit{`it may be rational to be irrational sometimes'}.

% irrational behaviour may be considered quasi-rational --- to act irrationally even though a player has no incentive (in the reward structure) for doing so

\section{New notion: $\alpha$-RNE with herding}

In classical game theory, it is assumed that all players are perfectly rational, and then, the widely-known and accepted Nash Equilibrium (NE) is provided as the solution of the game. However, in reality, we rarely encounter such perfectly rational players. Instead, more often than not, players take decisions based on some simple rules. The most common of such rules is the one where players exhibit herding behaviour; for example, in a stock market, players tend to buy the derivative that they believe the majority of the players will purchase, or on a traffic signal, people cross the road when they see others crossing the road, etc.

Our aim in this paper is to propose an appropriate notion of equilibrium that caters to such a mix of rational and irrational (to be more specific, the ones with the herding behaviour) players.  %It is easy to observe that the rational players have precedence in taking the actions, while the irrational players wait, observe and then choose their actions. In view of such chronology of play, we propose a new notion of equilibrium below.
Towards this, consider a large population and assume that there are $\alpha$ fraction of rational players, while the remaining population is composed of irrational players. Each player has to choose an action from the set of actions, denoted by ${\mathcal A}$, where $|{\mathcal A}| < \infty$. For each $a \in \mathcal{A}$, let $\mu(a)$ be the fraction of players who choose action $a$; define $\mu := (\mu(a))_{\{a \in \mathcal{A}\}}$. Similarly, let $\mu^R:= (\mu^R(a))_{\{a \in \mathcal{A}\}}$ be the empirical distribution corresponding to (only) rational players.  

Define the function $u : \mathcal{A} \times [0,1]^{|{\mathcal A}|} \to \mathbb{R}$ to represent the utility of players. 
Thus, each player receives the utility $u(a, \mu)$ if it chooses an action $a$ and the empirical distribution of the actions by the rest of the population is $\mu$. Note that the utility function is the same for all the players and depends upon $\mu$ like in mean-field games (\cite{carmona2018probabilistic}).  As in classical theory, the rational players are capable of performing extensive computations and thus, choose an action that maximizes their utility. Hence, if $\mu$ were the empirical distribution of the actions chosen by the entire population\footnote{Note that the game is described in the mean-field framework, therefore, the action chosen by a single player does not affect the outcome of the game (see \cite{carmona2018probabilistic}). Given this, it is appropriate to view $\mu$ as the empirical distribution corresponding to the `entire' population.},  the best response ($\mu^R$) of any rational player against $\mu$ would satisfy the following:
\vspace{2mm}

\hspace{-3mm}\fbox{
\begin{minipage}{0.45\textwidth}
\vspace{-2mm}
\begin{align}\label{eqn_NE_rational}
    \supp(\mu^R) \subseteq {\rm Arg} \max_{a\in {\mathcal A}} u(a, \mu),
\end{align}
\end{minipage}
}
\vspace{2mm}

where
\begin{align}\label{eqn_support}
    \supp(\mu) := \{a \in \mathcal{A}: \mu(a) > 0\}.
\end{align}
On the other hand, irrational players exhibit herding behaviour --- they blindly follow others and do not optimize like rational players. One simple way to model the herding behaviour of the irrational players is to assume that such players choose an action that is played the most by rational players, i.e., they choose the action $a$ which satisfies $\mu^R(a) \geq \mu^R(a')$ for all $a' \neq a$. 
However, since typically irrational players cannot distinguish rational from irrational players, we consider a more realistic way of capturing the herding behaviour of the irrational players. We assume that each irrational player chooses an action played by the majority among all other players, including irrational players. To be precise, we assume that each irrational player chooses the following action, against $\mu$:
\vspace{2mm}

\hspace{-3mm}\fbox{
\begin{minipage}{0.45\textwidth}
\vspace{-2mm}
\begin{align}\label{eqn_NE_irrational}
        f(\mu) &:= \min_i\left\{a_i \in {\rm Arg} \max_{a \in \mathcal{A}} \mu(a)\right\}.
\end{align}
\end{minipage}
}
\vspace{2mm}

In the above, for simplicity and tractability of the analysis, we assume that the action with the smallest index in the set ${\rm Arg} \max_{a \in \mathcal{A}} \mu(a)$ is preferred in case of a tie. Observe that the action chosen by the irrational players as per the above rule may not be the best response to $\mu$; it is just a response driven by the herding behaviour of the players.

Finally, the proportion of players choosing different actions in $\mathcal{A}$ is given by:
\vspace{2mm}

\hspace{-3mm}\fbox{
\begin{minipage}{0.45\textwidth}
\vspace{-2mm}
\begin{align}\label{eqn_NE_total}
    \mu(a) = \alpha \mu^R(a) + (1-\alpha) 1_{\left\{a = f(\mu)\right\}}, \mbox{ for each } a \in \mathcal{A}.
\end{align}
\end{minipage}
}
\vspace{2mm}

The above relationship is obtained as $\alpha \mu^R(a)$-fraction of rational players choose the action $a$, and all irrational players choose the same action $a$ only if $a = f(\mu)$ (see \eqref{eqn_NE_irrational}).

At this point, one should note that rational players choose an action anticipating the response of the irrational players (as $\mu$ depends on $f(\mu)$, see \eqref{eqn_NE_rational} and \eqref{eqn_NE_total}). Thus, we define a pair $(\mu, \mu^R)$ to be an equilibrium if it is satisfies:
\begin{enumerate}
    \item[(i)] $\mu^R \in \mbox{Best Response}(\mu)$; 
    \item[(ii)] the empirical measure ($\mu$) of the population is given by \eqref{eqn_NE_total}, when the corresponding counterpart for rational players is given by $\mu^R$; and
    \item[(iii)] $f(\mu)$ denotes the majority action as in \eqref{eqn_NE_irrational}.
\end{enumerate}
The above discussion is formally summarized below:
\begin{definition}\label{defn_alphaRNE}
    A pair of empirical measures $(\mu, \mu^R)$ is called an \underline{$\alpha$-Rational Nash Equilibrium}, or in short, $\alpha$-RNE, if it satisfies \eqref{eqn_NE_rational}, \eqref{eqn_NE_irrational} and \eqref{eqn_NE_total}.
\end{definition}

%%%% do not remove!
% It ``may not'' be possible to define an appropriate stackelberg game as the solution of the game is represented only by $\mu^R$, while the definition of $\alpha$-rne is in terms of a tuple.

Observe that the above definition is a natural extension of the NE defined in the classical mean-field games (MFGs), where `all' players are rational and optimize the utility function $u(\cdot; \mu)$ (see \cite{carmona2018probabilistic}). Thus, \eqref{eqn_NE_rational} is satisfied by $\mu$ in MFGs at the NE (not by $\mu^R$), i.e.,
\begin{align}\label{eqn_NE}
    \supp(\mu) \subseteq {\rm Arg} \max_{a\in {\mathcal A}} u(a, \mu).
\end{align}

Let us now immediately consider an example to see how $\alpha$-RNE differs from the classical NE.
\begin{figure*}
  \centering
  \medskip
\begin{subfigure}{.28\textwidth}
\centering
    \includegraphics[trim = {16cm 14cm 10cm 4cm}, clip, scale = 0.16]{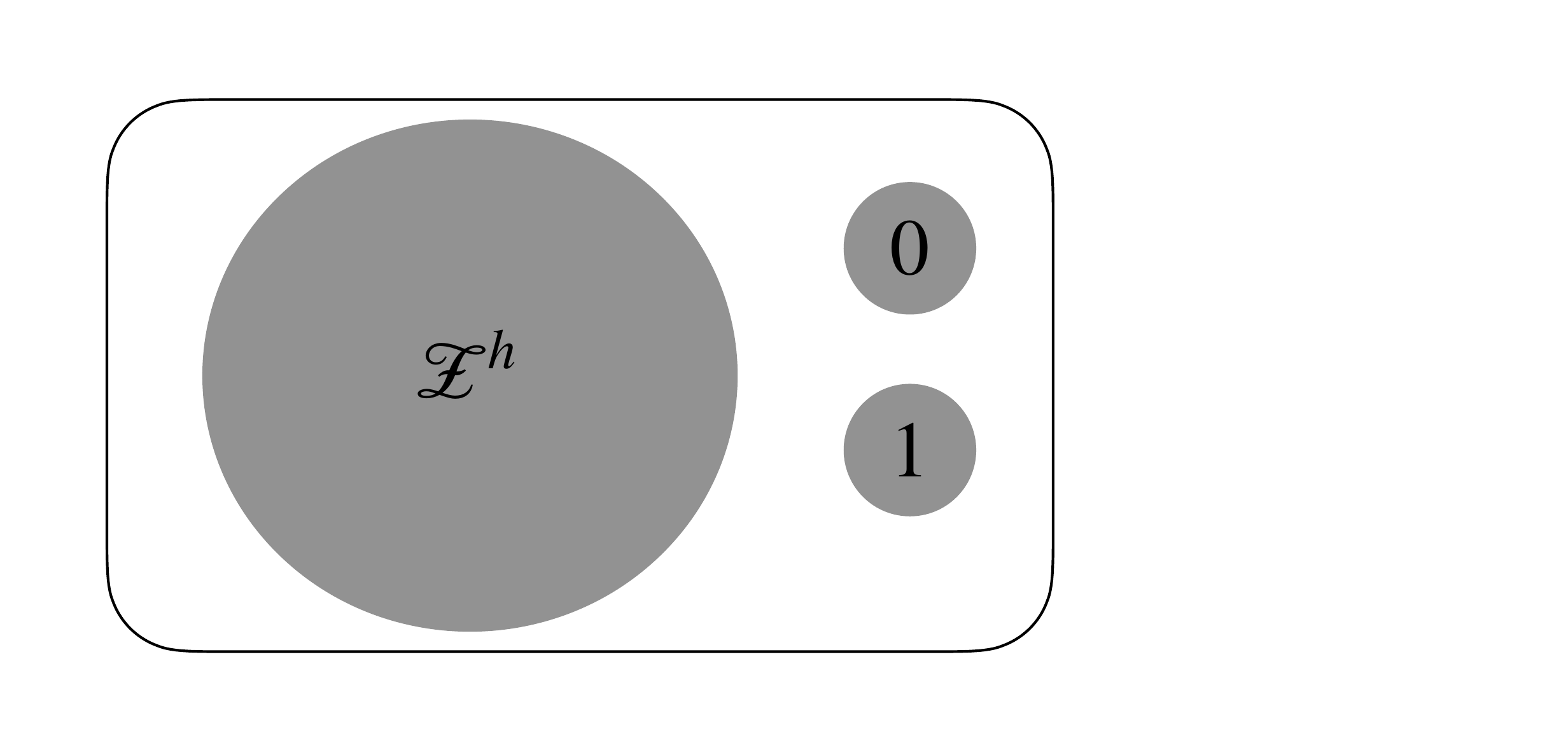}
    \caption{when $\alpha \in \left(\frac{1}{2}, 1\right]$}
\end{subfigure}
\begin{subfigure}{.34\textwidth}
\centering
    \includegraphics[trim = {16cm 14cm 3cm 4cm}, clip, scale = 0.16]{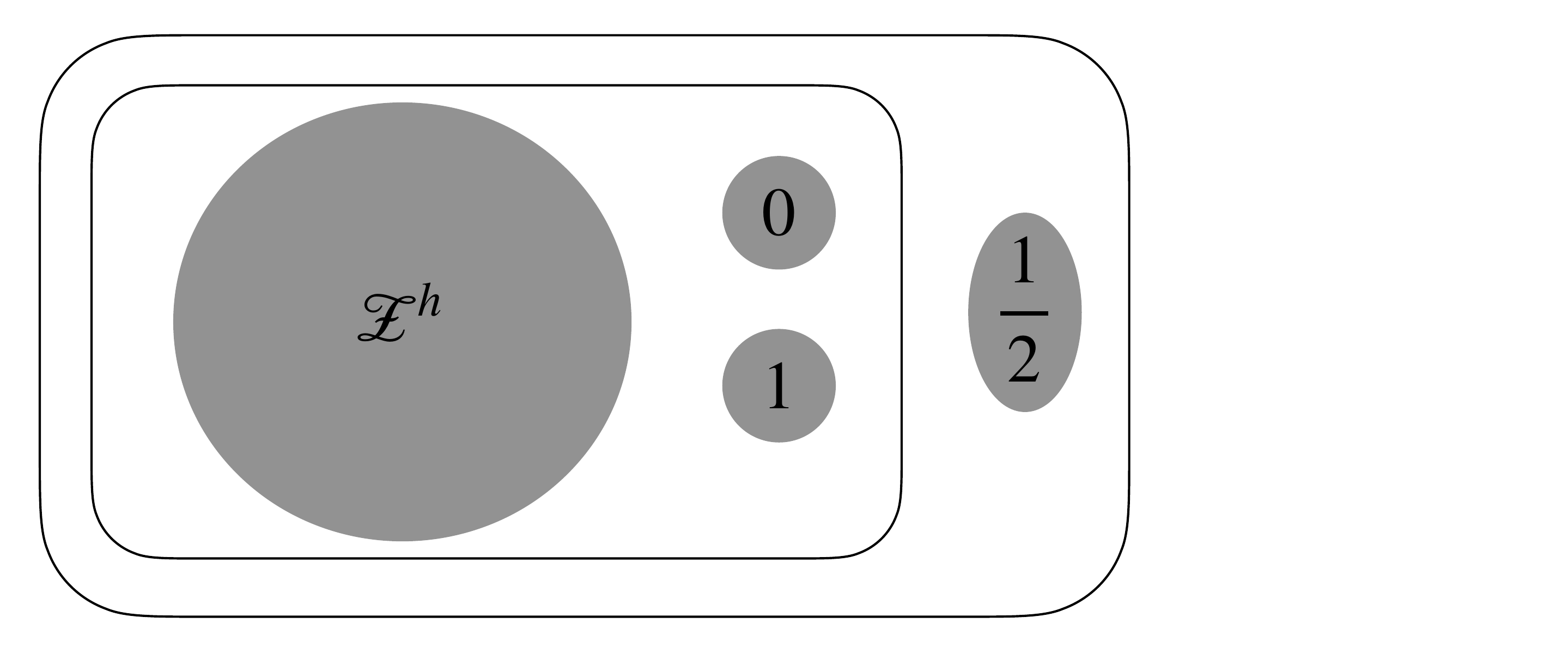}
    \caption{when $\alpha = \frac{1}{2}$}
\end{subfigure}
\begin{subfigure}{.34\textwidth}
\centering
    \includegraphics[trim = {16cm 14cm 13cm 4cm}, clip, scale = 0.16]{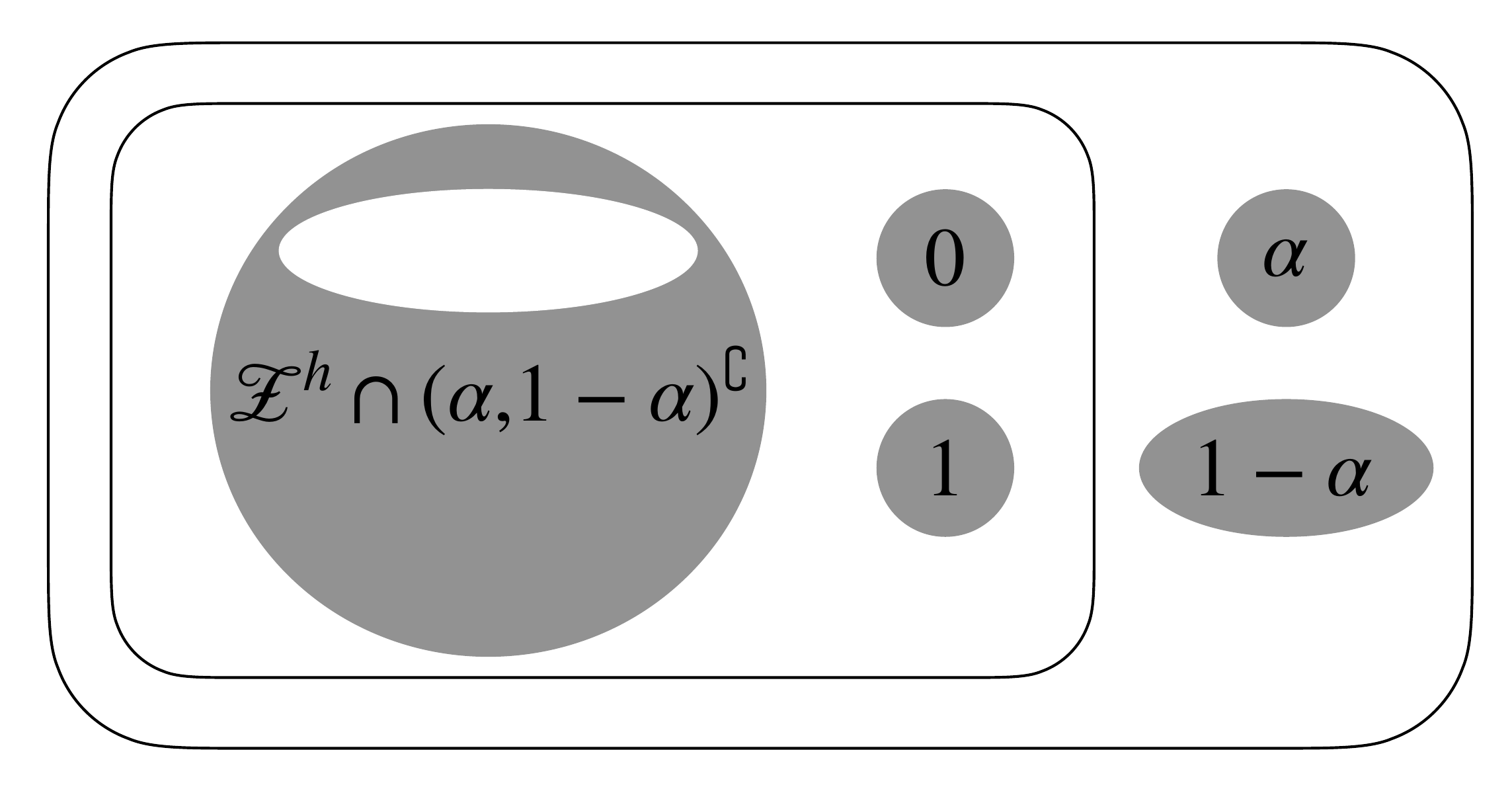}
    \caption{when $\alpha \in \left(0, \frac{1}{2}\right)$}
\end{subfigure}
\caption{Candidates for $\NE$ (shaded region) --- new equilibria emerge when $\alpha \leq \frac{1}{2}$, some equilibria are deleted when $\alpha < \frac{1}{2}$}
\label{fig:alpha_RNE}
\end{figure*}

\begin{example}
Consider a game with three actions, i.e., $\mathcal{A} = \{1,2,3\}$. Assume that the utility function $u(a, \mu)$ satisfy the following relation:
$$
u(1,\mu) > u(2,\mu) > u(3,\mu), \mbox{ for each } \mu.
$$
Clearly, ${\rm Arg} \max_{a \in \mathcal{A}}$ =$\{1\}$. Thus, for $\mu_1$ to be a classical NE, $\supp(\mu_1)$ should equal $\{1\}$, see \eqref{eqn_NE}. Hence, there is a unique classical NE,  $\mu_1 = (1,0,0)$, for the game.

Now, consider the realistic scenario where irrationals outnumber the rational players, i.e., $\alpha < \nicefrac{1}{2}$. Then, one can easily verify that $(\mu_\alpha,\mu^R_\alpha)$ with $\mu_\alpha = \mu_1$ as above and $\mu^R_\alpha = (1,0,0)$ is an $\alpha$-RNE. Further, $(\mu_\alpha,\mu^R_\alpha)$, with $\mu_\alpha = (\alpha,1-\alpha,0)$ and $\mu^R_\alpha$ as before is also an $\alpha$-RNE. There can be more $\alpha$-RNEs. 
\end{example}

The above example illustrates that an $\alpha$-RNE need not be a classical NE. We delve deep into the new notion for the game with two actions, where we derive more interesting insights. Further, we also derive the simple conditions to identify the $\alpha$-RNEs.

\section{Game with two actions}
% \begin{figure*}
%     \centering\includegraphics[trim={0cm 2cm 0cm 2cm}, clip, scale=0.25]{alphaRNE.pdf}
%     \caption{$\alpha$-RNEs, when $\mathcal{A} = \{1, 2\}$}
% \end{figure*} 

Let us consider that players can choose either action $1$ or $2$, i.e., ${\mathcal{A}} = \{1, 2\}$. For this setting, we propose simpler notations as follows: (i) let $z$ be the proportion of players who choose $a = 1$ and thus, $\mu = (z, 1-z)$ and (ii) let  $y$ be the proportion of rational players (among rational players) who choose $a = 1$ and thus,  $\mu^R = (y, 1-y)$. As a result, we write the utility function $u(\cdot, \mu)$ as $u(\cdot, z)$. 

Finally, the $\alpha$-RNE is given by $(\mu, \mu^R) \equiv (z, y)$ which satisfy the following:
\vspace{2mm}

\hspace{-3mm}\fbox{
\begin{minipage}{0.45\textwidth}
\vspace{-2mm}
\begin{align}
    z &= \alpha y + (1-\alpha) 1_{\left\{z \geq \frac{1}{2}\right\}} \mbox{ and} \label{eqn_alpha_RNE2_first}\\
    \supp(\mu^R) &\subseteq {\rm Arg} \max_{a\in {\mathcal A}} u(a, z). \label{eqn_alpha_RNE2_second}
\end{align} 
\end{minipage}
}
\vspace{2mm}

Note that in the underlying case with two actions, $f(\mu) = 1$ only if $z\geq \nicefrac{1}{2}$; therefore, \eqref{eqn_NE_irrational} and \eqref{eqn_NE_total} together leads to \eqref{eqn_alpha_RNE2_first} given above.

Now, by \eqref{eqn_alpha_RNE2_first},  $\alpha$-RNE can be represented only in terms of the proportion $z$. Denote the set of $\alpha$-RNEs by $\NE$. Then, one can easily verify that the set $\NE$ has the following structure:
\begin{align}\label{eqn_set_NE}
\begin{aligned}
        \NE &= \left\{z : (z, y^*(z)) \mbox{ is an $\alpha$-RNE}\right\}, \mbox{ where}\\
        y^*(z) &:= 
    \begin{cases}
        \frac{z}{\alpha}, &\mbox{ if } z < \frac{1}{2}, \\
        1-\frac{1-z}{\alpha}, &\mbox{ if } z \geq \frac{1}{2}.
    \end{cases}
    \end{aligned}
    \end{align}
In the above, $y^*(z)$ is provided by solving \eqref{eqn_alpha_RNE2_first}. Thus, by virtue of the above structure, it is sufficient to solve for \eqref{eqn_alpha_RNE2_second} alone, instead of solving for \eqref{eqn_alpha_RNE2_first} and \eqref{eqn_alpha_RNE2_second} simultaneously. Now, observe that ${\rm Arg}\max_{a\in \mathcal{A}} u(a, z)$ can be simply recognized by comparing the utilities $u(1, z)$ and $u(2, z)$ in the underlying case. This motivates us to define the following function:
\begin{align}\label{eq_fun_h}
    h(z) := u(1,z) - u(2,z).
\end{align}
The idea is to identify the $\alpha$-RNEs using the zeroes of the function $h$, i.e., from the set:
\begin{align}
    \zh := \left \{z^* \in [0,1] : h(z^*) = 0\right\}.
\end{align} 
% For example, suppose $z^* = \alpha \in \zh$ for some utility function, when $\alpha < \frac{1}{2}$. Then, $\supp(\mu^R) = \{1\} \subset \{1, 2\} = {\rm Arg} \max_{a\in {\mathcal A}} u(a, z)$, see \eqref{eqn_set_NE}. Thus, such a $z^* \in \NE$. We generalise this idea for any $z^* \in [0,1]$ and present it precisely below:
%

\subsection{Identification of Equilibria}
At first, for the sake of reference and comparisons, we provide the set of classical NEs.
\begin{theorem}[\textbf{Identification of MFG-NEs}]\label{prop}
    Suppose ${\mathcal{A}} = \{1, 2\}$. Then, the set of classical NEs:
 \begin{equation}
        \NEc \subseteq \zh \cup \{0, 1\} \label{prop11}.
    \end{equation} For the converse, we have:
    \begin{enumerate}
        \item[(i)] $\zh \subseteq \NEc$,
        \item[(ii)] $0 \in \NEc$ only if $h(0) \leq 0$, and
        \item[(iii)] $1 \in \NEc$ only if $h(1) \geq 0$.
    \end{enumerate}
\end{theorem}
\begin{proof}
    See Appendix \ref{appendix}.
\end{proof}
The above result provides simple conditions to identify classical NEs. It states that every zero of $h$ is a NE. Further, $0$ and $1$ can be NE also if $h(0) < 0$ and $h(1) > 0$ respectively. Next, we identify the set of $\alpha$-RNEs in terms of $\NEc$.
\hide{
\begin{theorem}[\textbf{Identification of $\alpha$-RNEs}]\label{thrm_alphaRNE_2}
    Suppose ${\mathcal{A}} = \{1, 2\}$. When $\alpha > \nicefrac{1}{2}$, $\NE = \NEc$. Otherwise, the following is true:
        \begin{align} \label{set_alphaRNE}
            \NE \subseteq
                \NEc \cup \{\alpha, 1-\alpha\} \backslash (\alpha, 1-\alpha).
        \end{align}
        For the converse of the above, we have:
        \begin{enumerate}
        \item[(i)] $\NEc \backslash (\alpha, 1-\alpha) \subseteq \NE$,
        \item[(ii)] $(1-\alpha) \in \NE$ if and only if $h(1-\alpha) \leq 0$, and 
        \item[(iii)] For $\alpha < \nicefrac{1}{2}$, $\alpha \in \NE$ if and only if $h(\alpha) \geq 0$.
    \end{enumerate}
\end{theorem}
\begin{proof}
    See Appendix \ref{appendix}.
\end{proof}
}

\begin{theorem}[\textbf{Identification of $\alpha$-RNEs}]\label{thrm_alphaRNE_2}
    Suppose ${\mathcal{A}} = \{1, 2\}$. Then, the following statements are true:
    \begin{enumerate}
        \item[i)] when $\alpha > \nicefrac{1}{2}$, $\NE = \NEc$, 
        \item[(ii)] when  $\alpha \le \nicefrac{1}{2}$, then:
        \begin{align} \label{set_alphaRNE}
            \NE \subseteq
                \NEc \cup \{\alpha, 1-\alpha\} \backslash (\alpha, 1-\alpha).
        \end{align}
        For the converse of the above, we have:
        \begin{enumerate}
        \item[(a)] $\NEc \backslash (\alpha, 1-\alpha) \subseteq \NE$,
        \item[(b)] $(1-\alpha) \in \NE$ if and only if $h(1-\alpha) \leq 0$, and 
        \item[(c)] For $\alpha < \nicefrac{1}{2}$, $\alpha \in \NE$ if and only if $h(\alpha) \geq 0$.
    \end{enumerate}
    \end{enumerate}
\end{theorem}
\begin{proof}
    See Appendix \ref{appendix}.
\end{proof}
Interestingly, the above Theorem asserts that \textit{the presence of irrational players have no effect on the set of equilibria ($\NE = \NEc$) when the rational players outnumber the irrational players} ($\alpha > \nicefrac{1}{2}$).

More interestingly, the situation drastically differs when $\alpha \leq \nicefrac{1}{2}$. Firstly, two new equilibria can arise, namely $\alpha$ and $1-\alpha$, see \eqref{set_alphaRNE}, under the conditions specified in (ii.b), and (ii.c) respectively. Secondly, not every zero of $h$ can be an $\alpha$-RNE --- the zeroes which are only in $[0, \alpha] \cup [1-\alpha, 1]$ are allowed. Thus, \textit{some classical NEs are deleted and new equilibria are added when the irrational players form the majority}.

We summarize the above implications in \autoref{fig:alpha_RNE}.

\subsection{Comparison of utilities}\label{sec_user}
In our framework, both rational and irrational players participate in the game. Therefore, first and foremost, one would like to know whether the utility of rational players diminishes due to the presence of irrational players. Subsequently, one might be interested in knowing if the irrational players suffer due to their herding behaviour, when compared with the utility they could have obtained if they were rational. 

Further, it is known that NE often results in players gaining lesser utility than the social optimal utility. So happens because the NE provides stability only against unilateral deviations. However, if multiple players deviate from NE, higher utility can be achieved. Considering this, it is plausible that players' utility may be closer to the social optimal utility at $\alpha$-RNE, than at NE, since irrational players collectively deviate from NE. If this anticipation holds, it suggests a rational inclination towards irrationality. We formally investigate all these aspects below.

Denote the expected utility of a rational player at $z_\alpha^* \in \NE$ by $\ura(z^*_\alpha)$, for $0 < \alpha \leq 1$, and observe:
\begin{align}\label{eqn_util_r}
   \ura(z^*_\alpha) &= y^*(z^*_{\alpha}) u(1, z^*_\alpha) + (1-y^*(z^*_\alpha)) u(2, z^*_\alpha).
\end{align}
In the above, $y^*(z^*_{\alpha})$ is the probability of a rational player choosing action $1$. 
Similarly, denote the expected utility for the irrational player by $\uira(z^*_\alpha)$, and note that it is given by:
\begin{align}\label{eqn_util_irr}
    \uira(z^*_\alpha) &= 1_{\left\{z^*_\alpha \geq \frac{1}{2}\right\}} u(1, z^*_\alpha) + \left(1-1_{\left\{z^*_\alpha \geq \frac{1}{2}\right\}}\right) u(2, z^*_\alpha).
\end{align}
Recall that our motive is to compare the utilities of the players under $\alpha$-RNE and classical NE. Keeping this in mind, we aim to investigate if, under $\alpha$-RNE, players can achieve utility that is comparable to the social optimal utility  ($\us$) under the classical setting ($\alpha = 1$). Thus, define:
\begin{align}\label{eqn_social}
    \us := \sup_{z \in [0,1]} \left(z u(1, z) + (1-z) u(2, z) \right).
\end{align}
Notice that the social optimal utility in our case\footnote{To be precise, the social optimal utility when $\alpha \in(0,1]$ is given by:
\vspace{-2mm}

{\tiny
\begin{align*}
\begin{aligned}
    \us_\alpha &:= \sup_{z \in [0,1]} \Bigg(\left\{\alpha y^*(z) + (1-\alpha)1_{\left\{z\geq \frac{1}{2}\right\}}\right\} u(1, z) \\
    &\hspace{15mm}+ \left\{\alpha(1-y^*(z)) + (1-\alpha)\left(1- 1_{\left\{z\geq \frac{1}{2}\right\}}\right)\right\} u(2, z)\Bigg).
\end{aligned}
\end{align*}}} is different than the one defined above. Thus, it is possible that the utility of a rational/irrational player can exceed $\us$ when $\alpha < 1$ --- if this happens, then it would mean that such  players are benefiting from the presence of herding-irrational players; in fact, this is true for the rational players in all the examples considered in Section \ref{sec_examples}.

% Now, suppose that there exists some dynamics that converge to $z_\alpha \in \NE$. Then, we say that:
% \begin{enumerate}
%     \item[P$_1$:] \textbf{it is rational to be irrational at $z_\alpha \in \NE$} \textit{if both rational and irrational players receive higher utility at $z_\alpha$ than the utility that (rational) players receive at any classical NE}, i.e., if:
% \end{enumerate}

% % Observe that if the above condition is true, then there would be no incentive for an irrational player to change their behaviour from herding to rational-type. 

We begin with comparing the utilities for a rational and irrational player under $\alpha$-RNEs. Further, we compare the utilities with the social optimal utility. 

\begin{proposition}\label{Proposition_util} For any $\alpha \in (0,1)$,  $\uira(z^*_\alpha) \leq \ura(z^*_\alpha)$ and $\uira(z^*_\alpha) \leq \us$, for all $z^*_\alpha \in \NE$. 
\end{proposition}
\begin{proof}
    See Appendix \ref{appendix}.
\end{proof}
The above result asserts that rational players always obtain more utility than irrational players. Additionally, the irrational players can never achieve higher utility than the social optimal utility ($\us$). 

Recall from Theorem \ref{thrm_alphaRNE_2} that some new equilibria may get added or classical NEs may be deleted in the presence of irrational players. The next result states that if new equilibria (namely, $\alpha$ and $1-\alpha$) are not added (i.e., $\NE \subseteq  \NEc$), then both the rational players and more importantly, the irrational players attain exactly as much as a rational player gets in the classical setting. 
\begin{proposition}\label{Proposition_util_equal}
    When $\NE \subseteq \NEc$, then $\us \geq \ura(z^*_\alpha) = \uira(z^*_\alpha) = \ur(z^*_\alpha)$ for all $z^*_\alpha \in \NE$.
\end{proposition}
\begin{proof}
    See Appendix \ref{appendix}.
\end{proof}
Thus, for example, if more rational players are present in the system ($\alpha > \nicefrac{1}{2}$), then they are able to manipulate the irrational players in such a way that no player loses anything. Further, by the above result, it is clear that no player (not even a rational player) receives more than the social optimal utility ($\us$), when $\NE \subseteq \NEc$.

Now, observe that the above result comments on all $\alpha$-RNEs, except $z^*_\alpha \in \{\alpha, 1-\alpha\}$ such that $z^*_\alpha \in \NE$ but $z^*_\alpha \not\in \zh$. Thus, if $\NEc \subseteq \NE$, then according to Proposition \ref{Proposition_util}, it is evident that the irrational players definitely receive strictly less utility than the rational players at $\alpha$-RNE. However, several interesting possibilities arise at $z^*_\alpha \in \{\alpha, 1-\alpha\}$ under said conditions:
\begin{itemize}
    \item [(i)] the rational players may outperform the best possible outcome in an all-rational case, i.e., $\ura(z^*_\alpha) > \us$. This holds for all the examples discussed in the subsequent section.
    \item[(ii)] the irrational players may experience no change or a loss ($\uira(z^*_\alpha) < \ur(z_1^*)$) or a gain ($\uira(z^*_\alpha) > \ur(z_1^*)$) when compared to some classical NE ($z_1^* \in \NEc$). The first two scenarios occur for the non-atomic routing game, while the latter applies to the other two games discussed in the coming section.
    \item[(iii)] interestingly, further, we shall see in the bandwidth sharing game that both rational and irrational players benefit at $\alpha$-RNE such that for all $z_1^* \in \NEc$:
    \begin{align*}
        \ura(z^*_\alpha) > \ur(z_1^*) \mbox{ and } \uira(z^*_\alpha) > \ur(z_1^*).
    \end{align*}
    Clearly, the utility of rational and irrational players surpasses and approaches closer to the social optimal utility ($u^S$) respectively. Thus, in such cases, we may declare that `it is rational to be irrational'.
\end{itemize}

\section{Examples}\label{sec_examples}

% {\color{red} check if any desired NE is deleted.}

\input{pigou}

\subsection{Participation Game} \label{sec_participation}
Motivated by \cite{agarwal2023single}, we consider the game where each player has to decide whether to participate or not in an activity. Let us denote $a = 1$ as the action indicating participation and $a=2$ as the action of non-participation in the activity. Thus, $z$ denotes the proportion of participants.

Each non-participant gets a (perceived) utility equal to $1$. To increase participation, the game designer provides a fixed utility $C < 1$ to each participant, and additionally, it offers a reward of $P > 0$, which is equally distributed among all the participants. Hence, the utility function can be expressed as follows: 
\begin{align} \label{utility_participation}
    u(a, z) = \left(C + \frac{P}{z}1_{\{z > 0\}}\right)1_{\{a = 1\}} + 1_{\{a = 2\}}. 
\end{align}
For the above game, the set of $\alpha$-RNEs depends on the value of $P$. We consider two disjoint regimes: (i) $P\geq 1-C$ and (ii) $P < 1-C$, and present the results for the respective regimes below. The proofs again follow by Theorem \ref{thrm_alphaRNE_2}.
\begin{corollary}\label{cor_participation1}
    Consider the participation game with $P \geq 1- C$. Then, we have:
    \begin{enumerate}
        \item[(i)] when $\alpha \in [\nicefrac{1}{2}, 1]$, $\NE = \NEc = \{0,1\}$, and 
        \item[(ii)] when $\alpha \in (0, \nicefrac{1}{2})$, $\NE = \NEc \cup \{\alpha\}$. \eop
    \end{enumerate}
\end{corollary}
Under classical NE, dichotomy occurs: either everyone participates or no one participates. This situation may be undesirable for the designer as the chances of zero participation are $50\%$. Interestingly, when irrational players also play the game and constitute the majority, the designer can exploit the inherent herding behavior of irrationals and possibly induce $\alpha$-level of participation. Thus, the likelihood of non-zero participation increases due to herding.

Further, the rational players benefit when $\alpha$ is the $\alpha$-RNE, as they receive higher\footnote{Given $P \geq 1-C$. Therefore, $\Delta:=\frac{P}{1-C} > 1$. Now, observe   $\ur(\alpha) - \us = C - 1  - P + \frac{P}{\alpha} > 0$, only if $\alpha < \frac{1}{1+\frac{1-C}{P}} = \frac{\Delta}{1+\Delta}$. Since $\alpha \in \NE$ when $\alpha < \frac{1}{2}$, and $\frac{\Delta}{1+\Delta} > \frac{1}{2}$, therefore, $\ur(\alpha) > \us$.} utility ($\ura(\alpha) = C+\nicefrac{P}{\alpha}$) than the social optimal utility ($\us = 1+P$), see \eqref{eqn_social}.

% Based on Theorem \ref{thrm_util}, the following utility comparisons arise:
% \begin{align*}
%     \ura(0) &= \uira(0) = \ur(0) = 1 < \us = 1+R, \\
%     \ura(1) &= \uira(1) = \ur(1) = C+R < \us, \mbox{ and}\\
%    \uira(\alpha) &= 1 < \ura(\alpha) = C+\frac{R}{\alpha}.
% \end{align*}
% It is clear that , $\uira(\alpha) < \us$, but the interesting thing to note here, $\us < \ura(\alpha)$.

\begin{corollary} \label{cor_participation2}
     Consider the participation game with $P < 1- C$. Then, we have:
    \begin{enumerate}
        \item when $\alpha \in (\nicefrac{1}{2}, 1]$, $\NE = \NEc = \left\{0,1, \frac{P}{1-C}\right\}$, 
        \item when $\alpha = \nicefrac{1}{2}$, 
        $$
            \NE = 
            \begin{cases}
                \NEc \cup \left\{\frac{1}{2}\right\}, &\mbox{ if } P < \frac{1-C}{2}, \\
                \NEc, &\mbox{ otherwise, and}
            \end{cases}
        $$ 
        \item when $\alpha \in (0, \nicefrac{1}{2})$, define $P_1 := \alpha(1-C)$ and $P_2 := (1-\alpha) (1-C)$. Then:
        $$
            \hspace{3mm} \NE = 
            \begin{cases}
                \NEc \cup \{1-\alpha\}, &\mbox{ if } 0 < P \leq P_1, \\
                \{0,1, \alpha, 1-\alpha\}, &\mbox{ if } P_1 < P < P_2, \\
                \NEc \cup \{\alpha\},  &\mbox{ if }  P_2 \leq P < 1-C. \hspace{3mm}\mbox{\eop}
            \end{cases}
        $$ 
    \end{enumerate}
\end{corollary}
As observed before, here also, either new equilibria are added, or some classical NEs are deleted when $\alpha \leq \nicefrac{1}{2}$. As one can anticipate,  lesser rewards imply lesser utility for the players. When $P < 1-C$, rational players get lesser utility (at $\alpha, 1-\alpha$) than the social optimal utility; recall previously, $\ura(\alpha) > \us$. Interestingly, $\ur(1) < \uira(z_\alpha^*)$ for $z_\alpha^* \in \{\alpha, 1-\alpha\}$. This suggests that the performance can potentially be improved if irrational players are involved. 

% diff. between social optima and irr. utility = R, 1-C, when R >= 1-C ---- min diff. = 1-c

% difff...... = R, R, 1-C, R, 1+R-(C + R/1-alpha)

% \begin{align*}
%     \us &= 1+R, \\
%     \mbox{with more R }, \ur(\alpha) &= C+\frac{R}{\alpha} > \us, \uira(\alpha) = 1\\
%     \mbox{with less R }, \ur(\alpha) &= C+\frac{R}{\alpha} < \us, \uira(\alpha) = 1, \\
%     \ur(1-\alpha) &= 1 < \us,  \uira(1-\alpha) = C + \frac{R}{1-\alpha} < \uira(\alpha).
% \end{align*}

However, the most interesting fact about this game is that \textit{lesser reward means higher participation} under the realistic setting with $\alpha < \nicefrac{1}{2}$. To be precise, $(1-\alpha)$-level (more than $50\%$) of participation can occur when $P < P_2$ (also compare it with Corollary \ref{cor_participation1}). This happens because rational players tend to lose interest in participation with low reward, so they choose $a = 2$; irrational players (which form the majority) then choose $a = 1$ due to herding. 

In all, less reward proves detrimental for the players but advantageous for the game designer.

\subsection{Bandwidth Sharing Game}\label{sec_bandwidth}
Consider a communication network where players share the bandwidth to transmit their signals/information. The players can either transmit at the maximum capacity (which equals $1$) of the shared channel, or they can transmit at a lower level, which equals $\nicefrac {1}{2}$. We refer to the two actions as $a=1$ and $a=2$ respectively. In the first case, the communication of others can get interfered with, while in the latter case, no disruption occurs. 
The overall utility derived by any player depends upon its maximum capacity discounted by the overall interference caused by the opponents, and hence, the utility function is as follows:
\begin{eqnarray} \label{Eqn_communication_game}
u(a,z) = \left ( 1_{\{a=1\}} +  \frac{1}{2}  1_{\{a=2\}} \right ) (1-z).
%1_{\{a=1\}} + \left(\frac{1-z}{2}\right)1_{\{a=2\}}
\end{eqnarray}
The above game is a simplified version of the bandwidth sharing game discussed in \cite{narahari2014game} for classical strategic form setup; here we have also modified it for the mean-field setting.
Now, we will provide the set of $\alpha$-RNEs for this game, which can be derived by Theorem \ref{thrm_alphaRNE_2}.
\begin{corollary}\label{Bandwidth_corollary}
    Consider the bandwidth sharing game. The set of $\alpha$-RNEs is as below:
    \begin{itemize}
       \item[(i)] when $\alpha \in [\nicefrac{1}{2},1]$,  $\NEc = \NE =  \{1\}$, and
       \item [(ii)] when $\alpha \in (0,\nicefrac{1}{2})$, $\NE = \NEc \cup \{\alpha\}$.\eop
    \end{itemize} 
\end{corollary}
It is easy to solve the social optimization problem \eqref{eqn_social} for this game:  the social optimal utility $\us$ equals $\nicefrac{1}{2}$, which is realized when no one transmits at capacity $1$. 

  When there are only rational players (i.e.,  $\alpha = 1$),   clearly from \eqref{Eqn_communication_game}, the  unique classical NE equals $z^*_1 = 1$  and  the corresponding utility $\ur(1) =0$. Now, consider $\alpha < \nicefrac{1}{2}$.  
  Then, by \eqref{eqn_util_r}, \eqref{eqn_util_irr}, one can calculate that $\ura(\alpha) = 1-\alpha$ and $\uira(\alpha) = \nicefrac{1}{2}\left(1-\alpha\right)$.

  Interestingly, the rational players have strictly improved their utility compared to the scenario with all rational players, as 
  $\ur(1) < \ura(\alpha)$.  
%
 % 
 % This implies that as irrational players join, the utility of rational players increases compared to their utility under classical NE (i.e., $\ur(1) < \ura(\alpha)$).
More 
 interestingly, the utility of irrational players is also higher than that of the rational players under classical NE, as  $\ur(1) < \uira(\alpha)$.  
 \textit{Thus, in this case, it is `rational' to be irrational!} 
 
 Moreover, as $\alpha$ approaches $0$, the utility of  the irrational players  $\nicefrac{1}{2} \left(1-\alpha\right)$  approaches $\us = \nicefrac{1}{2}$. This is a surprising outcome --- at a selfish equilibrium, the players are achieving near social optimal utilities. Notably, the existence of a small fraction of rational players and a large fraction of herding players achieves this feat. 
 
As said before, $\ur(z_1^*) \leq \us $ for all $z_1^* \in \NEc$. However, here, rational players achieve more utility than the social optimal utility as $\us < \ura(\alpha)$ due to the presence of irrational players. Conclusively, the introduction of irrationality can be beneficial in some cases.

We summarize the main observations obtained for the three examples discussed in this section below. 

\begin{table}[htbp]
\centering
\renewcommand{\arraystretch}{1.7} % Adjust the stretch factor as needed
\resizebox{8.5cm}{!} {
\begin{tabular}{|l|ccc|}
\hline
\multicolumn{1}{|c|}{\large{Performance at $\alpha$-RNEs}}                           & \multicolumn{3}{c|}{\large{Name of the game}}                                                                       \\ \hline
\multicolumn{1}{|c|}{}                           & \multicolumn{1}{c|}{\large{Non-atomic routing}} & \multicolumn{1}{c|}{\large{Participation}}  & \large{Bandwidth Sharing }          \\ \hline
\large{$u^R_\alpha \geq u^I_\alpha$}           & \multicolumn{1}{c|}{\large{yes}}                & \multicolumn{1}{c|}{\large{yes}}            & \large{yes}                         \\ \hline
 \large{$u^R_\alpha(z_\alpha^*) > \us$}                  & \multicolumn{1}{c|}{\large{yes ($z_\alpha^* = \alpha$)}}  & \multicolumn{1}{c|}{\large{yes ($z_\alpha^* = \alpha$, higher $P$)}}               &   \large{yes ($z_\alpha^* = \alpha$)}                           \\ \hline
 \large{$u^I_\alpha(z_\alpha^*) > u^R_1(z_1^*)$ for some $z_1^* \in \NEc$} & \multicolumn{1}{c|}{\large{no}}                 & \multicolumn{1}{c|}{\large{yes ($z_\alpha^* \in \{\alpha, 1-\alpha\}$, lower $P$)}} & \multicolumn{1}{c|}{\large{yes ($z_\alpha^* = \alpha$)}} \\ \hline
\large{rational to be irrational}                      & \multicolumn{1}{c|}{\large{no}}                 & \multicolumn{1}{c|}{\large{no}}             & \large{yes}                         \\ \hline
\end{tabular}}
\caption{Highlights of the examples}
\end{table}

\section{Conclusions}
This paper studies the mean-field game involving $\alpha$-fraction of rational and $(1-\alpha)$-fraction of irrational players. While rational players adhere to classical game theory principles, irrational players exhibit herding behavior by blindly choosing the action chosen by the majority. We introduce a novel equilibrium concept, termed $\alpha$-Rational Nash equilibrium ($\alpha$-RNE), which extends the NE for the classical mean-field games by capturing the responses of the irrational players.

For the games with two actions, our findings reveal that the presence of irrational players can alter the set of equilibria compared to classical NEs. New equilibria may emerge, while some classical equilibria may disappear when more irrational players are in the system. Otherwise, the set of equilibria does not change. We also provide easy conditions to identify $\alpha$-RNEs for such games.

Under $\alpha$-RNE, a rational player never obtains lower utility than an irrational player. Moreover, the utility for irrational players never surpasses the social optimal utility achievable when all players are rational. We show through examples that the presence of irrational players may benefit rational players, leading to them achieving more than the social optimal utility for certain $\alpha$-RNEs. Importantly, our study suggests that it is sometimes `rational' for the players to behave irrationally.  

% In fact, the rational players under $\alpha$-RNE may sometimes achieve more than the social optimal utility possible when all the players are rational. Additionally, there are examples where irrational players achieve higher utility than rational players (under classical NE). Such examples illustrate that it is sometimes rational for players to behave irrationally.

\textbf{Future directions:} 
One of our future objectives is to formulate an appropriate Stackelberg game involving rational and herding-irrational players and to compare it with the mean-field game with herding introduced in this paper. 

It will also be worthwhile to identify the learning dynamics which converge to the $\alpha$-RNEs and to investigate the same in the context of behavioural game theory. 

Reflecting on our previous findings in the participation game, wherein the presence of irrational players proved advantageous for the game designer, we are motivated to explore the mechanism design aspect in the presence of herding-irrational players.

% To illustrate this point, consider a scenario where students have to choose between pursuing a career in (say) science or commerce. Some students make their choices based on their capabilities, while others simply follow prevailing trends (herding behavior). In such instances, it becomes imperative for institutions to estimate the proportion of students selecting science or commerce due to genuine interest (i.e., $y^*(z), 1-y^*(z)$) and identify the option driven by herding behavior. Such a study is important because if some option is chosen merely due to herding, then the advancement of that field will be effected.

% Towards this, the values of $y^*(z)$ are easy to determine using \eqref{eqn_set_NE}, for any $z \in \NE$. Further, if $z = \alpha$ or $1-\alpha$, then either action $2$ or $1$ is chosen only by irrational players\footnote{ To verify this, say $z = \alpha$; then $y^*(\alpha) = 1$, as $\alpha \in \NE$ only when $\alpha < \frac{1}{2}$ (see \eqref{eqn_set_NE}).}, respectively. In such a case, the said actions under respective $\alpha$-RNEs are not sustainable. For example, if science is chosen only due to herding, then research in science will be affected. 

\bibliographystyle{IEEEtran}
\bibliography{references}

% for arxiv

\input{appendix}
\hide{\subsection{Minority Game}
Let us first consider a simple game involving two restaurants (denoted as $1$ and $2$) in a town. Each player has to decide which restaurant to go to (i.e., $a = 1$ or $2$). Players have a preference for less crowded restaurants. Thus, each player receives a utility of $1$ if they choose the less crowded restaurant and $0$ otherwise. This description results in the following utility function:
\begin{align}\label{eqn_util_minority}
    u(a, z) = 
    \begin{cases}
        1, &\mbox{ if } z \leq \frac{1}{2} \mbox{ and } a = 1, \\
        1, &\mbox{ if } z \geq \frac{1}{2} \mbox{ and } a = 2, \\
        0, &\mbox{ otherwise}.
    \end{cases}
\end{align}
If all players were rational, the obvious best strategy would be to distribute equally among the two restaurants (i.e., $z = \nicefrac{1}{2}$), resulting in an equal utility of $1$ for each player. However, with the presence of irrational players, rational players must adapt their strategy to divert the attention of irrational players in order to still enjoy a less crowded restaurant.
Our concept of $\alpha$-RNE accommodates games with such contradictory behaviors among players. We determine the $\alpha$-RNEs for the underlying game in the subsequent result (proof directly follows from Theorem \ref{thrm_alphaRNE_2}).
\begin{corollary}
    For the minority game, we have:
    \begin{enumerate}
        \item[(i)] when $\alpha \in [\nicefrac{1}{2}, 1]$, $\NE = \NEc = \left\{\nicefrac{1}{2}\right\}$, and 
        \item[(ii)] when $\alpha \in (0,\nicefrac{1}{2})$, $\NE = \left\{\alpha, 1-\alpha\right\}$.  \eop
    \end{enumerate}
\end{corollary}
The above corollary suggests that if rational players constitute more than $50\%$ of the population, they need not alter their strategy. In fact, in this scenario, both rational and irrational players achieve utility equal to the social optimal utility ($= 1$).

Further, note that when irrationals are present and $\alpha < \nicefrac{1}{2}$, then the classical NE is removed and two new equilibria ($\alpha, 1-\alpha$) emerge (see Theorem \ref{thrm_alphaRNE_2}). Under these equilibria, rational players select a restaurant, and then, driven by intrinsic behavior, irrational players choose the other restaurant. Here, one can observe how \textit{herding behavior detrimentally affects irrationals}, leading to a utility of $0$, compared to the utility of $1$ gained by the rational players.

% \begin{align*}
%     \ura(z) &= \us = 1 >  \uira(z) = 0, \mbox{ for } z \in \{\alpha, 1-\alpha\}\\
%     \ura\left(\frac{1}{2}\right) &= 1 = \uira\left(\frac{1}{2}\right) = \us = \ur\left(\frac{1}{2}\right).
% \end{align*}
}

% \hide{\input{additional_calculations}}
\end{document}

%% file: pigou.tex
\subsection{Non-atomic Routing Game}\label{sec_pigou}
Consider a non-atomic routing game, which is a slight modification of Pigou's network game (\cite{narahari2014game}). One can travel from source (S) to destination (T) via hub $1$ or $2$. The users must opt for either the path via hub $1$ or hub $2$ to minimize their travel time.

It takes $\gamma z$ minutes to travel from S to T, for some $\gamma \in (1, \infty)$, while the travel time via hub $2$ is just $1$ minute. Consider that $\alpha$-fraction of users are rational, and the rest adhere to the majority's choice (as in \eqref{eqn_NE_irrational}). Rational users base their decisions on optimizing the following utility function:
\begin{align}\label{eqn_util_braess}
    u(a, z) = \left(-\gamma z\right)1_{\{a = 1\}} + \left(-1\right)1_{\{a = 2\}}, 
\end{align}where $z$ is the proportion of users travelling via hub~$1$.

Next, we provide the set of $\alpha$-RNEs for the above game. The proof is omitted as it directly follows from Theorem \ref{thrm_alphaRNE_2}.
\begin{corollary} \label{pigou's-corollary}
Consider the non-atomic routing game and define $\Delta:= \nicefrac{1}{\gamma}$. Then, the set of $\alpha$-RNEs is given below in two regimes:
\begin{enumerate}
    \item[(i)] when $\Delta \leq \frac{1}{2}$:
    $$
    \NE = 
    \begin{cases}
        \{\alpha, 1-\alpha\}, &\mbox{ if } \alpha \leq \Delta, \\
        \{\Delta, 1-\alpha\}, &\mbox{ if } \Delta < \alpha \leq \frac{1}{2}, \\
        \{\Delta\}, &\mbox{ if } \alpha > \frac{1}{2};
    \end{cases}
    $$
    % \item[(ii)] $\Delta = \frac{1}{2}$:
    % $$
    % \NE = 
    % \begin{cases}
    %     \{\alpha, 1-\alpha\}, &\mbox{ if } \alpha < \frac{1}{2}, \\
    %     \{\Delta\}, &\mbox{ if } \alpha \geq \frac{1}{2}.
    % \end{cases}
    % $$
    \item[(ii)] when $\Delta > \frac{1}{2}$: 
    $$
    \hspace{1cm}\NE = 
    \begin{cases}
        \{\alpha, 1-\alpha\}, &\mbox{ if } \alpha \leq 1-\Delta, \\
        % \{\Delta, \alpha, 1-\alpha\}, &\mbox{ if } \alpha = 1-\Delta, \\
        \{\Delta, \alpha\}, &\mbox{ if } 1-\Delta < \alpha < \frac{1}{2}, \\
        \{\Delta\}, &\mbox{ if } \alpha \geq \frac{1}{2}. \hspace{1.5cm} \mbox{\eop}
    \end{cases}
    $$
\end{enumerate}
\end{corollary}
At first, note that the classical NE is unique and equals $\Delta$. A new equilibrium emerges when $\alpha > \min\{\Delta, 1-\Delta\}$. Otherwise, the classical NE is removed, and two new equilibria emerge.

Thus, under classical setting ($\alpha = 1$), $\Delta$-fraction of users choose to travel through hub $1$. However, if we consider the game with the rational and irrational users, then the congestion on the path via hub $1$ can either remain the same (as before), or it can decrease to $\alpha$-level, or increase to $(1-\alpha)$-level. Even at times when congestion is lesser, the irrational users are at a loss in this game as\footnote{Here, $\uira(z_\alpha^*) = -1$ for $z_\alpha^* \in \{\Delta, \alpha\}$. Further, $\uira(1-\alpha) = \nicefrac{-(1-\alpha)}{\Delta} < \ur(\Delta)$; to verify this, note the conditions when $1-\alpha \in \NE$ from Corollary \ref{pigou's-corollary}.} $\uira(z^*_\alpha) \leq \ur(\Delta)$, for all $z^*_\alpha \in \NE$. 

But interestingly, the \textit{rational users exploit the presence of irrational users and benefit at multiple levels:}

(i) rational users take an equal or less amount of time to travel, compared to irrational users (as $\ura(z^*_\alpha) \geq \uira(z^*_\alpha)$ for any $z^*_\alpha \in \NE$, by Proposition \ref{Proposition_util}).

(ii) \textit{rational users take lesser or equal amount of time to travel than under all-rational case} (as\footnote{Here, $\ur(\Delta) = -1 = \ura(z_\alpha^*)$ for $z_\alpha^* \in \{\Delta, 1-\alpha\}$. Further, $\ura(\alpha) = \nicefrac{-\alpha}{\Delta} > \ur(\Delta)$; to verify this, note the conditions when $\alpha \in \NE$ from Corollary \ref{pigou's-corollary}.} $\ura(z^*_\alpha) \geq \ur(\Delta)$, for all $z^*_\alpha \in \NE$). 

(iii) the utility  $\ura(\alpha)$ is strictly increasing in $\alpha$. Thus, if $\alpha$ emerges as the $\alpha$-RNE, then more irrational users in the system mean more advantage to the rational users. In fact, if $\alpha < \left(1-\nicefrac{\Delta}{4}\right)\Delta$, then the rational users receive more utility at $\alpha$ than the social optimal utility ($\us = \nicefrac{\Delta}{4}-1$).

%Here, the rational users greatly benefit due to less congestion on the path via hub $1$ induced due to the irrational users.

% Now, consider the system parameter $\gamma < 2$ (i.e., $\Delta > \frac{1}{2}$) and the realistic setting with $\alpha< \frac{1}{2}$. Then, by Theorem \ref{thrm_util}, $\ura(\alpha) > \ura(z^*)$ for $z^* \in \{\Delta, 1-\alpha\}$. Consequently, rational users take the least amount of time to reach the destination if `all' rational users travel through hub $1$ (i.e., when $\alpha$ is the equilibrium). Now, further if $\gamma < \frac{1}{1-\alpha}$, then both equilibria $\Delta$ and $\alpha$ favor the rational users as then they experience either less or equal time to reach the destination compared to classical NE (see Corollary \ref{pigou's-corollary}(ii), and again note that $\ura(\alpha) > \ura(\Delta)$).

% also, $\uira(\alpha) = \ur(\Delta)$. Further, $\uira(1-\alpha) = \frac{-(1-\alpha)}{\Delta}$. {\color{red}$\uira(1-\alpha) - \ur(\Delta) = 1 - \frac{1-\alpha}{\Delta} = \frac{\alpha - (1-\Delta)}{\Delta} < 0$.} So, it is not rational to be irrational.

%% file: appendix.tex
\section{Appendix}\label{appendix}

\hide{
\noindent \textbf{Proof of Theorem \ref{thrm_alphaRNE}:}
    Let $(\mu, \mu^R)$ be an $\alpha$-RNE. By Definition \ref{defn_alphaRNE}, $(\mu, \mu^R)$ satisfies \eqref{eqn_set_O}-\eqref{eqn_NE_rational}. By \eqref{eqn_NE_FTC}, it is clear that all irrational players choose the action from the set $\mathcal{O}$ such that its index is minimal; thus, condition (i) of the claim is true. Further, rational players choose an action which maximizes the utility, i.e., the one in \eqref{eqn_NE_rational}. This implies that condition (ii) of the claim is also true. 

    Conversely, given $(\mu, \mu^R)$ which satisfies (i) and (ii), let us establish the characterisation of $\mathcal{O}$. 
    
    Let $a^* := f^*(\mu)$. Then, by (i), we get:
    \begin{align*}
        \mu(a^*) &= \alpha \mu^R(a^*) + (1-\alpha), \\
        \mu(a) &= \alpha \mu^R(a), \mbox{ for all } a \in \mathcal{A} - \{a^*\}.
    \end{align*}
    Thus, $a^* \in \mathcal{O}$.

    For any $a \neq a^*$, we have:
    $$
    \alpha \mu^R(a^*) + (1-\alpha) = \mu(a^*) \geq \mu(a) = \alpha \mu^R(a).
    $$

    Consider any $a \neq a^*$ which is not in ${\rm Arg} \max \mu(a)$. Then, we aim to prove that $a \not \in \mathcal{O}$. 
    \begin{align*}
        (1-\alpha) + \alpha\mu^R(a) &= 1-\alpha + \mu(a) \\
        &< 1-\alpha + \mu(a^*) \\
        &=  1-\alpha + 1-\alpha + \alpha \mu^R(a^*)
    \end{align*}

    By definition of $f^*(\mu)$, $\mu(f^*(\mu)) = \mu(a) > \mu(a')$ for all $a \in {\rm Arg} \max \mu(a)$ and $a' \in \mathcal{A} - {\rm Arg} \max \mu(a)$.
    Thus:
    \begin{align*}
        \alpha \mu^R(f^*(\mu)) + (1-\alpha) &= \mu(f^*(\mu)) = \mu(a) = \alpha \mu^R(a), a \in {\rm Arg} \max \mu(a) \\
        \alpha \mu^R(f^*(\mu)) + (1-\alpha) &= \mu(f^*(\mu)) > \mu(a) = \alpha \mu^R(a), a \in \mathcal{A} - {\rm Arg} \max \mu(a)
    \end{align*}
    
    This implies that $\mu^f(f^*(\mu); \mu^R) = 1$, thus satisfying \eqref{eqn_NE_FTC}. 

    {\color{red}tie breaking}
    
    By definition of $\supp(\mu^R)$, $\mu^R(a) > 0$ for all $a \in \supp(\mu^R)$ and $\mu^R(a) = 0$ for all $a \in \mathcal{A}-\supp(\mu^R)$. Let $u^* := u(a, \mu)$ for some $a \in \supp(\mu^R)$. Then, by (ii), $u(a, \mu) = u^*$ for all $a \in \supp(\mu^R)$ and $u^* \ge u(a', \mu)$ for all $a' \in \mathcal{A}-\supp(\mu^R)$. Thus, the non-zero proportion $\mu^R(a)$ is only for those actions $a$ which have the maximum utility, i.e., in Arg max of \eqref{eqn_NE_rational}. Conclusively, by Definition \ref{defn_alphaRNE}, $(\mu, \mu^R)$ is an $\alpha$-RNE. \eop
    
\vspace{2mm}}

\newcommand{\Au}{{\mathcal A}_u}

\noindent \textbf{Note:} \textit{Only in this section, we refer ${\rm Arg}\max_{a \in \mathcal{A}} u(a, z)$ as $\Au$, in short.}

\vspace{2mm}

\noindent \textbf{Proof of Theorem \ref{prop}:}     Before we start the proof, note that when $\alpha = 1$, \eqref{eqn_alpha_RNE2_first} implies $z = y$. By \eqref{eqn_alpha_RNE2_second}, for any $z$ to be in $\NEc$, following should hold:
    \begin{align}\label{eqn_NE_z}
        \supp(\mu) \subseteq \Au, \mbox{ for } \mu = (z, 1-z).
    \end{align}
    Now, consider any $z \in \NEc$. Then, three different cases arise based on the value of $h(z)$. Suppose $h(z) > 0$. By \eqref{eq_fun_h}, $\Au = \{1\}$. Under \eqref{eqn_NE_z}, $\supp(\mu) = \{1\}$ (since $z \in \NEc$); thus, $z = 1$. Similarly, one can show that $z = 0$ if $h(z) < 0$. Lastly, suppose $h(z) = 0$. Then, $\Au = \{1, 2\}$, again by \eqref{eq_fun_h}. Since $z\in \NEc$, therefore, under \eqref{eqn_NE_z}, either $\supp(\mu) = \{1\}$ or $\{2\}$ or $\{1, 2\}$. In the first two cases, $z = 1, 0$ respectively as above. In the last case, $z \in (0,1)$.  Combining all the implications from above, we get \eqref{prop11}.  
    
    We now prove the claims for the converse of \eqref{prop11}. Suppose $z \in \zh \cap (0,1)$. Then, $\supp(\mu) = \{1, 2\} = \Au$. Thus, $z \in \NEc$. Next, suppose $z = 0$. Then, $\supp(\mu) = \{2\}$. If $h(0) \leq 0$, then $\Au = \{1, 2\}$ or $\{2\}$; thus, $0 \in \NEc$. However, if $h(0) > 0$, then $\Au = \{1\}$. Clearly, it contradicts \eqref{eqn_NE_z} and thus, $0 \not\in \NEc$ in this case. One can similarly prove the claim for $z = 1$. Conclusively, (i)-(iii) hold. \eop

\vspace{2mm}

\noindent \textbf{Proof of Theorem \ref{thrm_alphaRNE_2}:} We divide the proof into two cases.

\noindent \textbf{(i)} when $\alpha > \frac{1}{2}$
\vspace{2mm}

\noindent \textbf{Claim 1:} $\NE \subseteq \NEc$

Suppose $z \in \NE$. Firstly, let $h(z)>0$. By \eqref{eq_fun_h}, $\Au=\{1\}$. Under \eqref{eqn_alpha_RNE2_second}, $\supp(\mu^R) = \{1\}$; thus, $y^*(z) = 1$. By \eqref{eqn_set_NE}, $z = 1$ (as $\alpha > \frac{1}{2}$). Thus, $z = 1 \in  \NEc$, as $\{1\} \subseteq \NEc$ (see Theorem \ref{prop}). Similarly, one can show that $z = 0 \in \NEc$ when $h(z) < 0$. Further, if $h(z) = 0$, then, $z \in \zh$. By Theorem \ref{prop}(i), it is clear that $z \in \NEc$.
In all, $\NE \subseteq \NEc$.

% Secondly, let $z \in \NE$ such that $h(z) < 0$. Then $\Au = \{2\}$, by \eqref{eq_fun_h}. Under \eqref{eqn_alpha_RNE2_second}, $\supp(\mu^R) = \{2\}$, which implies $y^*(z)= 0$. By \eqref{eqn_set_NE}, $z = 0$. Here, also, by the above logic, $z \in  \NEc.$

\vspace{2mm}

\noindent \textbf{Claim 2:} $\NEc \subseteq \NE$

Let $z \in \NEc$. Then by \eqref{prop11}, either $z \in \zh$ or $z \in \{0,1\}$. Say $z \in \zh \cap \left[0, \frac{1}{2}\right)$. Here, $y^*(z) = \frac{z}{\alpha}$; observe $y^*(z) = 0$ if $z = 0$ and $y^*(z) \in (0, 1)$ otherwise. Thus, $\supp(\mu^R) = \{2\}$  or $\{1, 2\} \subseteq \Au$, as  $z \in \zh$. By Definition \ref{defn_alphaRNE}, $z \in \NE$. One can  prove in a similar manner that any $z \in \zha \cap \left[\frac{1}{2}, 1\right)$ is also in $\NE$.

Now, say $z = 0$. Observe $y^*(0) = 0$. Thus, $\supp(\mu^R) = \{2\}$. Recall $z = 0 \in \NEc$ only if $h(0) \leq 0$, by Theorem \ref{prop}(ii). Thus, $\Au = \{1, 2\}$ or $\{2\}$. Then, as above, $z \in \NE$. The proof similarly follows when $z = 1$. Hence, $\NEc \subseteq \NE$. 

Conclusively, $\NEc = \NE$. 

\vspace{2mm}
\noindent \textbf{(ii)} when $\alpha \leq \frac{1}{2}$ 
\vspace{2mm}

\noindent \textbf{Claim 1:} \eqref{set_alphaRNE} holds

Say $z \in \NE$. Then, $z$ satisfies $\eqref{eqn_alpha_RNE2_second}$. Now, we divide the proof based on the values of $h(z)$. 

Firstly, let $h(z)>0$. Then, as in the case with $\alpha > \frac{1}{2}$, $y^*(z) = 1$. By \eqref{eqn_set_NE}, either $z = \alpha$ if $\alpha< \frac{1}{2}$ or $z = 1$. Observe that $\alpha \not\in \NEc$ and $\{1\} \subseteq \NEc$. This implies that $z \in  \NEc \cup \{\alpha, 1-\alpha\} \backslash (\alpha, 1-\alpha).$
Similarly, one can prove \eqref{set_alphaRNE} when $h(z) < 0$. At last, let $h(z) = 0$. Then, $\Au = \{1, 2\}$, again by \eqref{eq_fun_h}. Under \eqref{eqn_alpha_RNE2_second}, three possibilities arise:
    \begin{itemize}
      \item $\supp(\mu^R) = \{1\}$: here, $z = \alpha$ if $\alpha< \frac{1}{2}$ or $z = 1$, as when $h(z) > 0$.
        \item $\supp(\mu^R) = \{2\}$: here,  $z = 1-\alpha$ if $\alpha \leq \frac{1}{2}$ or $z = 0$, as when $h(z) < 0$.
        \item $\supp(\mu^R) = \{1, 2\}$: here, $y^* \in (0,1)$. This implies that either $z < \alpha$ and $z < \frac{1}{2}$, or $z > 1-\alpha$ and $z \geq \frac{1}{2}$. 
    \end{itemize}
    In all three cases, one can easily see that $z \in  \NEc \cup \{\alpha, 1-\alpha\} \backslash (\alpha, 1-\alpha)$. Conclusively, \eqref{set_alphaRNE} holds.

    \vspace{2mm}
    
\noindent \textbf{Claim 2:} statements (a), (b) and (c) hold

    (a) Suppose $z \in \NEc \backslash (\alpha, 1-\alpha)$. Then by \eqref{prop11}, $z \in \zh\backslash (\alpha, 1-\alpha)$ or $z \in \{0,1\}$. 
    
    Firstly, say $z \in \zh\backslash (\alpha, 1-\alpha)$ such that $z < \frac{1}{2}$. Here, $y^*(z) = \frac{z}{\alpha}$; observe $y^*(z) = 1$ if $z = \alpha$ and $y^*(z) < 1$ otherwise. Thus, $\supp(\mu^R) = \{1\}$ or $\{1, 2\}$; in either case, $\supp(\mu^R) \subseteq \Au$, as  $z \in \zh$. This implies that $z$ satisfies \eqref{eqn_alpha_RNE2_second}. By Definition \ref{defn_alphaRNE}, $z \in \NE$. One can  prove in a similar manner that any $z \in \zh \cap \left[\frac{1}{2}, 1\right) \backslash (\alpha, 1-\alpha)$ is also in $\NE$.

    When $z = 0$ or $1$, the claim holds exactly as in the case with $\alpha > \nicefrac{1}{2}$.

(b) Let $\alpha < \nicefrac{1}{2}$. Say $\alpha \in \NE$. Then, $y^*(\alpha) = 1$, and hence $\supp(\mu^R) = \{1\} \subseteq \Au$ only if $h(\alpha) \geq 0$. Conversely, say $h(\alpha) \geq 0$. Then, $\Au = \{1, 2\}$ or $\{1\}$. It is now easy to observe that $\alpha \in \NE$ as $y^*(\alpha) = 1$ ensures \eqref{eqn_alpha_RNE2_second} is satisfied.

    (c) The proof follows as in part (ii.c). \eop
    \hide{Suppose $\alpha \leq \frac{1}{2}$. Then, we have:
    \begin{align*}
        1-\alpha \in \NE &\iff y^*(\alpha) = 0 \\
        &\iff \mu^R = (0, 1) \\
        &\iff \supp(\mu^R) = \{2\} \subseteq \Au\\
        &\iff h(\alpha) \leq 0.
    \end{align*}
    }
    
\hide{
\vspace{2mm}

\noindent \textbf{Proof of Theorem \ref{thrm_alphaRNE_2}:}
    Let $z \in \NE$. Then, we consider three different cases based on the value of $h(z)$.
    
    Firstly, suppose $h(z) > 0$. This implies that $\Au = \{1\}$, by \eqref{eq_fun_h}. Under \eqref{eqn_alpha_RNE2_second}, $\supp(\mu^R) = \{1\}$; thus, $y^*(z) = 1$. By \eqref{eqn_set_NE}, either $z = \alpha$ if $\alpha< \frac{1}{2}$ or $z = 1$.

    Secondly, consider $z$ such that $h(z) < 0$. Then $\Au = \{2\}$, by \eqref{eq_fun_h}. Under \eqref{eqn_alpha_RNE2_second}, $\supp(\mu^R) = \{2\}$, which implies $y^*(z))= 0$. By \eqref{eqn_set_NE}, either $z = 1-\alpha$ if $\alpha \leq \frac{1}{2}$ or $z = 0$.

    Lastly, suppose $h(z) = 0$. Then, $\Au = \{1, 2\}$, again by \eqref{eq_fun_h}. Since $z\in \NE$, therefore, under \eqref{eqn_alpha_RNE2_second}, three possibilities arise:
    \begin{enumerate}
      \item[(i)] $\supp(\mu^R) = \{1\}$: here, $z = \alpha$ if $\alpha< \frac{1}{2}$ or $z = 1$, as in first case above.
        \item[(ii)]  $\supp(\mu^R) = \{2\}$: here,  $z = 1-\alpha$ if $\alpha \leq \frac{1}{2}$ or $z = 0$, as in second case above. 
        \item[(iii)] $\supp(\mu^R) = \{1, 2\}$: here, $y^* \in (0,1)$. This implies that either $z < \alpha$ and $z < \frac{1}{2}$, or $z > 1-\alpha$ and $z \geq \frac{1}{2}$. 
    \end{enumerate}
      Combining all the implications from above, we get \eqref{set_alphaRNE}. We now prove the claims for the converse of \eqref{set_alphaRNE}. 
    
    (i) Claim: $\zha \subseteq \NE$. Below, we prove the claim for $\alpha \geq \frac{1}{2}$; the proof will follow analogously if 
    $\alpha < \frac{1}{2}$.
    
    Consider $z \in \mathcal{Z}_h^{(\alpha)} = \zh$ such that $z < \frac{1}{2}$. Here, $y^*(z) = \frac{z}{\alpha}$ and it clearly satisfies  \eqref{eqn_alpha_RNE2_first}. For $\mu^R = (y^*, 1-y^*)$, observe that $\supp(\mu^R) = \{1, 2\} \subseteq \Au$, as  $z \in \zh$. This implies that $(z, y^*(z))$ satisfies \eqref{eqn_alpha_RNE2_first}-\eqref{eqn_alpha_RNE2_second}, and thus by Definition \ref{defn_alphaRNE}, $z \in \NE$. One can  prove in a similar manner that any $z \in \zha \cap \left[\frac{1}{2}, 1\right)$ is also in $\NE$. 

    (ii) Consider $z = 0$. Observe $y^*(0) = 0$ satisfies \eqref{eqn_alpha_RNE2_first}. Thus, $\mu^R = (0,1)$ and $\supp(\mu^R) = \{2\}$. Here, if $h(0) \leq 0$, then $\Au = \{1, 2\}$ or $\{2\}$. Clearly then,  \eqref{eqn_alpha_RNE2_second} is satisfied. However, if $h(0) > 0$, then $\Au = \{1\}$, which contradicts \eqref{eqn_alpha_RNE2_second}. Therefore, $z \in \NE$ only if $h(0) \leq 0$.

    (iv) Consider $z = \alpha < \frac{1}{2}$ and $h(\alpha) > 0$. Then, $y^*(\alpha) = 1$ satisfies \eqref{eqn_alpha_RNE2_first}. Further, $\mu^R = (1,0)$ and $\supp(\mu^R) = \{1\} = \Au$. Thus, \eqref{eqn_alpha_RNE2_second} is also satisfied. This proves the claim. 

     One can easily complete the proof for part (iii) as in part (ii), and for part (v) as in part (iv). \eop
}

\vspace{2mm}

\noindent \textbf{Proof of Proposition \eqref{Proposition_util}:} 
Firstly, consider $z^* \in \NE \cap \zh$. Then, by definition of $\zh$, $u(1, z^*) = u(2, z^*)$. Thus, note from \eqref{eqn_util_r}, \eqref{eqn_util_irr} that:
\begin{align*}
    \ura(z^*) &= u(1, z^*) = \uira(z^*).
\end{align*}

Secondly, consider $z^* = 0 \in \NE \backslash \zh$. Then, $y^*(0) = 0$, and thus:
\begin{align*}
    \ura(0) = u(2, 0) = \uira(0).
\end{align*}
One can similarly prove that $\ura(1) = u(1, 1) = \uira(1)$, when $1 \in \NE \backslash \zh$.

Next, suppose $z^* = \alpha \in \NE \backslash \zh$. Then, by Theorem \ref{thrm_alphaRNE_2}, $h(\alpha) > 0$, i.e., $u(1, \alpha) > u(2, \alpha)$, and $\alpha < \frac{1}{2}$. The latter implies that $y^*(\alpha) = 1$; thus:
\begin{align*}
    \ura(\alpha) = u(1, \alpha) &> u(2, \alpha) = \uira(\alpha).
\end{align*}
Lastly, one can similarly prove the claim for $1-\alpha \in \NE \backslash \zh$.

Define $f(z) := z u(1, z) + (1-z) u(2, z) $ for all $z \in [0,1]$. Consider any $z^* \in \NE$, then (see \eqref{eqn_social}):
\begin{align*}
    \us &= \sup_{z \in [0,1]} f(z)\\
    &= \max\left\{\sup_{z \in [0,1]-\{z^*\}} f(z), f(z^*)\right\} \geq f(z^*) \geq  \uira(z^*).
\end{align*}
This completes the proof. \eop
\vspace{2mm}

\noindent \textbf{Proof of Proposition \eqref{Proposition_util_equal}:} 
Suppose $\alpha > \frac{1}{2}$. By Theorem \ref{thrm_alphaRNE_2}, $\NE = \NEc$. Further, $\us \geq \ura(z^*) = \uira(z^*)$, for each $z^* \in \NE$, see proof of Proposition \ref{Proposition_util}. From \eqref{eqn_util_r}:
\begin{align*}
    \ur(z^*) = 
    \begin{cases}
        u(2, z^*), \mbox{ if } z^* = 0 \in \NE \backslash \zh, \\
        u(1, z^*), \mbox{ otherwise}.
    \end{cases}
\end{align*}
Then, observe $\ur(z^*) = \uira(z^*)$, again from the proof of Proposition \ref{Proposition_util}. This completes the proof for $\alpha > \frac{1}{2}$.

Next, consider $\alpha \leq \frac{1}{2}$. Then, $\NE \subseteq \NEc$ only if $\alpha, 1-\alpha \not\in \NE$, see Theorem \ref{thrm_alphaRNE_2}. Thus, the proof follows as in case with $\alpha > \frac{1}{2}$.
 \eop

    % Lastly, consider $\alpha < \frac{1}{2}$. Here, $\NE \subseteq \NEc$ only if $\alpha, 1-\alpha \not\in \NE$. This again reduces to the case with $\alpha > \frac{1}{2}$, and thus the proof follows as before. \eop

%% file: main.bbl
\begin{thebibliography}{9}
\bibitem{narahari2014game} Narahari, Y. (2014). Game theory and mechanism design (Vol. 4). World Scientific.
\bibitem{camerer2011behavioral}Camerer, C. F. (2011). Behavioral game theory: Experiments in strategic interaction. Princeton university press.
\bibitem{thaler2018cashews} Thaler, R. H. (2018). From cashews to nudges: The evolution of behavioral economics. American Economic Review, 108(6), 1265-1287.
\bibitem{schultz2008introduction} Schultz, W. (2008). Introduction. Neuroeconomics: the promise and the profit. Philosophical Transactions of the Royal Society B: Biological Sciences, 363(1511), 3767-3769.
\bibitem{banerjee1992simple}Banerjee, A. V. (1992). A simple model of herd behavior. The quarterly journal of economics, 107(3), 797-817.
\bibitem{eliaz2002fault}Eliaz, K. (2002). Fault tolerant implementation. The Review of Economic Studies, 69(3), 589-610.
\bibitem{vasal2020alpha} Vasal, D., \& Berry, R. (2020). Alpha-Robust Equilibrium in Anonymous Games. Available at SSRN 3643821.
\bibitem{vasal2020fault} Vasal, D., \& Berry, R. (2020). Fault Tolerant Equilibria in Anonymous Games: best response correspondences and fixed points. arXiv preprint arXiv:2005.06812.
\bibitem{carmona2018probabilistic} Carmona, R., \& Delarue, F. (2018). Probabilistic theory of mean field games with applications I-II. Berlin: Springer Nature.
\bibitem{agarwal2023single} Agarwal, K., \& Kavitha, V. (2023, May). Single-out fake posts: participation game and its design. In 2023 American Control Conference (ACC) (pp. 2344-2350). IEEE.
\end{thebibliography}
